\documentstyle[12pt,aps,epsf]{revtex}
\vspace{2cm}
\begin{document}
\title{\bf PERSISTENT CURRENTS IN INTERACTING ELECTRONIC SYSTEMS}
\author{Georges Bouzerar and Didier Poilblanc}
\address{ 
Groupe de Physique Th\'eorique,
Laboratoire de Physique Quantique,\\
Universit\'{e} Paul Sabatier,
31062 Toulouse, France
}
\maketitle
\thispagestyle{empty}
\bigskip
{\small 
Persistent currents in disordered mesoscopic rings
threaded by a 
magnetic flux are calculated 
 using exact diagonalization methods
in the one-dimensional (1D) case and
self-consistent 
Hartree-Fock treatments for two dimensional (2D) systems.
For multichannel systems, a comparative study between
models of spinless or spinfull (Hubbard) fermions has been done. 
First, it is shown that a purely one-dimensional model
can not reproduce the correct order of magnitude of
the observed currents.
For 2D systems, going beyond first order pertubative calculations,
we show that the second harmonic of the current is
{\it strongly suppressed} in the case of spinless fermion models
but {\it significantly enhanced} for the Hubbard model.
This reduction (resp. increase) of the second harmonic
is related to a strong increase (resp. reduction) 
of the spacial charge density fluctuations.
Our work underlines the important role of the
spin degrees of freedom in the persistent currents. 
} 
\bigskip

\section{Introduction}

The observations of mesoscopic currents in very pure metallic 
nano-structures was done in pioneering experiments \cite{Levy,Chandra,Mailly}. 
In the first case, the experiment dealt with the average current 
of a system of $10^{7}$ disconnected rings in the diffusive regime while,
 in the second, a single ring was used. 
Although the existence of such persistent currents in small metallic rings
was predicted long ago \cite{history1,Buttiker}, the magnitude
of the observed currents
is still a real challenge to theorists.
Since studies neglecting electron-electron interaction
do not reproduce the order of magnitude of the currents \cite{gilles0,Altshuler,Schmid,Von Oppen},
there is a general belief that the interaction plays a crucial 
role in enhancing the current.
It has been suggested that the currents
could be enhanced to values close to the non disordered system
case \cite{Eckern,continuum}.
However, the exact role of the interaction in disordered systems is still 
unclear since treating interaction and disorder on equal footings is 
a difficult task.


\section{1D-Systems}

Let us first consider the simplest case of
pure 1D systems of spinless fermions.
The hamiltonian of the model reads,
\begin{eqnarray}
{\cal H}= -t/2  \sum_{< ij >} (exp(i2\pi \Phi/L) c_{\bf i}^{\dagger} 
c_{\bf j} +h.c.)
+V \sum_{< ij >}  n_{\bf i} n_{\bf j}  + \sum_{i} w_{\bf i} n_{\bf i}
\end{eqnarray}
where L is the size of the system and $\Phi= \frac{\phi}{\phi_{0}}$, 
where $\phi$ is the flux through the ring and $\phi_{0}=\frac{hc}{e}$.
The third term describes the disorder, 
where $n_{\bf i}= c_{\bf i}^{\dagger}c_{\bf i}$ and 
$w_{\bf i}$ are on site potentiel chosen randomly
in $[-W/2, W/2]$.
The nearest neighbour interaction has an amplitude V.

\begin{figure}[bth]                                                        
\begin{center}                                                                
{\parbox[t]{6cm}{\epsfxsize 6cm                                               
\epsffile{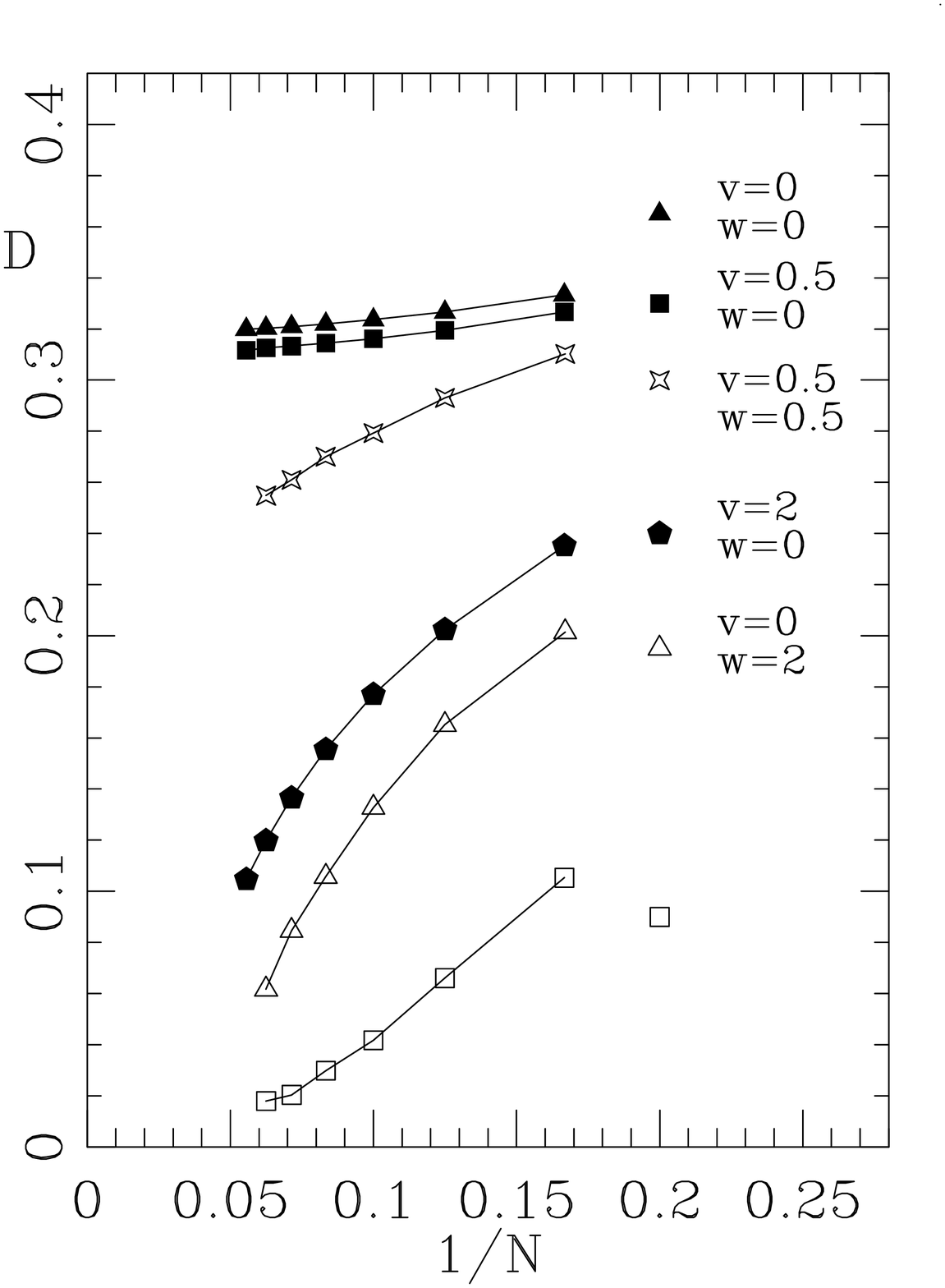} }                                                        \parbox[t]{6cm}{\epsfxsize 6cm                                                
\epsffile{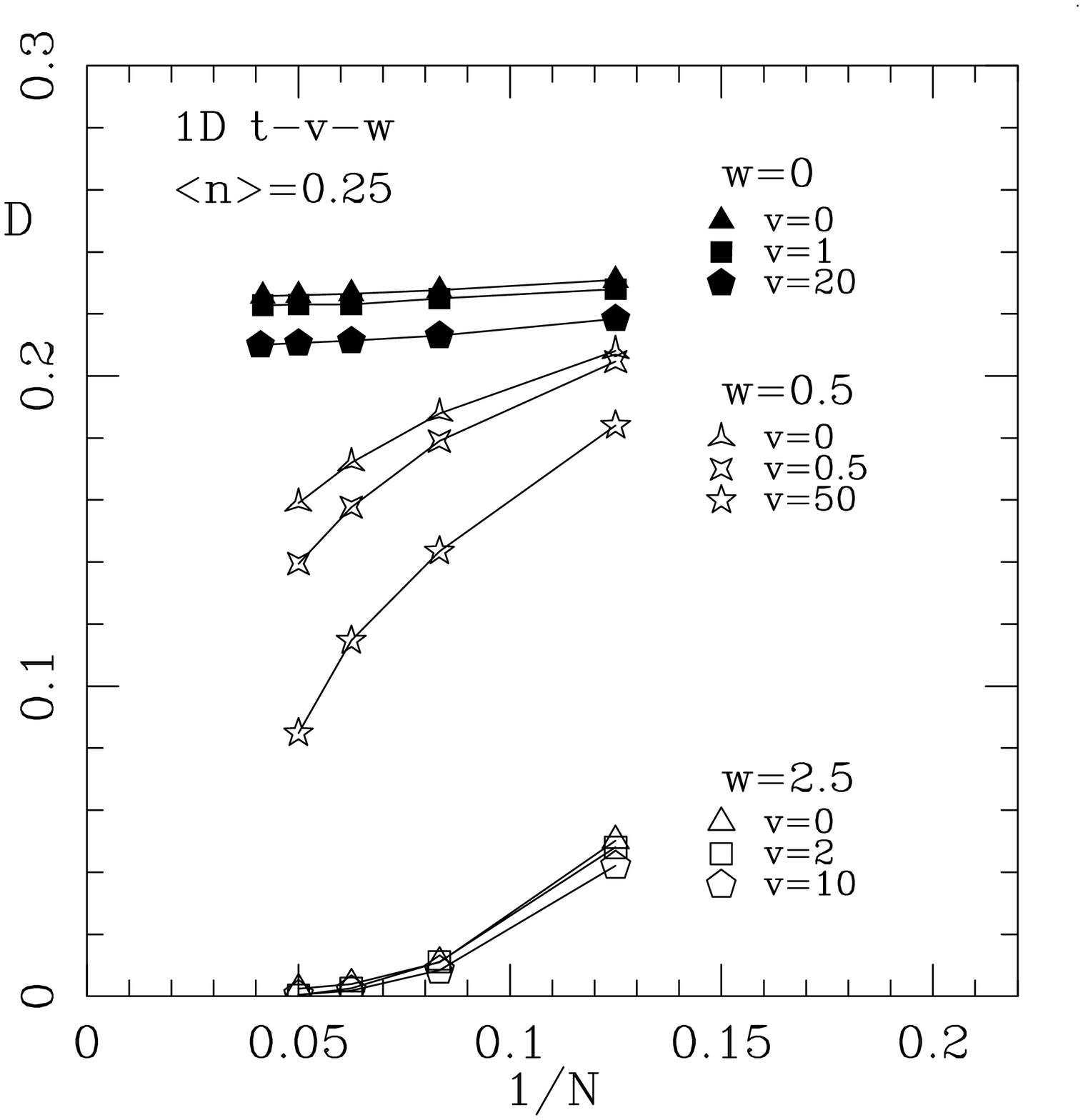}}                                                      
}   
\caption{Scaling of the Drude Weight
 at half-filling
(right) and quarter-filling (left) vs inverse system size.
The values of V and W (in units of t) are indicated on the picture.
Each dot represents an average value over at least
300 configurations of the disorder.}
\label{fig1}         
\end{center}
\end{figure}

\begin{figure}
\centerline{\epsfxsize 6cm \epsffile{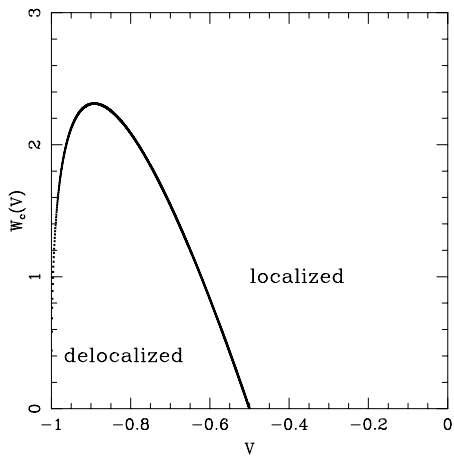}}
\caption{Phase diagram $W_{c}(V)$ vs V obtained from an RG approach.}
\label{fig2}
\end{figure}

\begin{figure}
\centerline{\epsfxsize 6cm \epsffile{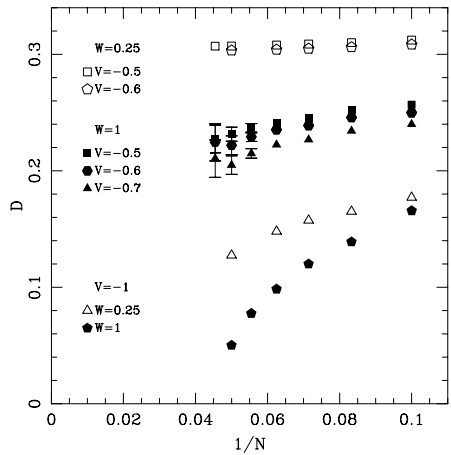}}
\caption{ED results of the Drude weight vs 1/N for $V < 0$.
Each dot represents an average value over at least
300 configurations of the disorder.
}
\label{fig3}
\end{figure}

If $E(\Phi)$ is the ground-state energy the persistent current
is given by,
\begin{eqnarray}
I=-\frac{1}{2\pi}\frac{\partial E}{\partial \Phi}  .
\end{eqnarray}
An alternative to quantitatively estimate the persistent current
is to calculate the Drude weight $D$,
\begin{eqnarray}
D=\frac{L}{4\pi^{2}} \big[ \frac{\partial^{2} E}
{\partial \Phi^{2}} \big]_{\Phi_{m}}  .
\end{eqnarray}
$\Phi_{m}$ corresponds to the minimum of $E (\Phi)$.
Using exact diagonalizations (ED) of small clusters
by the Lanczos method \cite{lanczos},
we have shown that, for strictly 1D systems \cite{meso1d}
of spinless fermions, the effect of a {\it repulsive} interaction is 
to increase further the localization of the
electrons and hence to decrease the value of the current.
In fig.\ref{fig1} we have plotted the Drude weight
as a function of the inverse system size for two different
band-fillings.
At half filling and for $W=0$, the Mott transition can
be easily seen from this picture.
The transition from metallic to insulator occurs at
$V_{c} = t$ (2t is the bandwidth).
Away from half-filling and for $W=0$, the system is
always metallic and the Drude weight depends  weakly on V.
When disorder is switched on, for increasing repulsive interaction, we
observe that D is reduced and, hence, the localization of the particles
is increased.
For $V > t$, the localization is both due
to the Anderson localization and Umklapp processes generated by V.
Note that, away from half-filling, we observe that the effect of
$V$ is very weak since Umklapp processes exist only in
higher orders of the interaction.
These results suggest that, for 1D systems of spinless
fermions, the interaction do not counteract the effect of
the disorder but, on the contrary, increases 
localization. Using a Hartree-Fock approach,
Kato et al. \cite{note_1d} have obtained qualitatively good agreement
with our ED results.
The case of {\it attractive} interaction
is generically very different. 
Indeed, for some values of the interaction parameters,
the persistent currents can be increased, even in the presence of disorder.
For the sake of simplicity, we shall concentrate here on the half-field case. 
At fixed V, the system
exhibits a Kosterlitz-Thouless-like transition \cite{KT} at a critical value
of the disorder parameter $W_{c}(V)$. 
In fig.\ref{fig2} we have plotted the phase diagram, calculated
within a renormalization group (RG) approach, in agreement with ref. \cite{DS}.
This picture suggests that
for $-1 < \frac{V}{t} < -0.5$ the system is metallic
for sufficiently weak disorder. Note that $W_{c}^{max} \simeq 2.3t$.
For $\frac{V}{t}<-1$ and any filling,
the system phase separates but this is of no interest for us 
in the present study.
We have compared the ED and the RG results \cite{meso1d2,RZ} 
and found a qualitatively good agreement.
In fig.\ref{fig3} the Drude weight is plotted
as a function of the inverse system size for various parameters V and W.
For weak disorder $ W < t $ the agreement is clear.
However, our numerical data suggest that the delocalized region is
restricted to a smaller part of the phase diagram.
This minor discrepancy might be due to the fact that the RG 
approach neglects Umklapp processes and is perturbative in V.
The difference between the repulsive and attractive
cases can be understood in the following way:
in the case of repulsive interaction the 
$2k_{f}$-fluctuations of the charge density are dominant.
Therefore, the pinning by the impurity potential is strong and leads to
a decrease of the localization length and, hence, to the reduction of the 
persistent currents.
On the other hand, in the attractive case,
the superconducting fluctuations are dominant.

For spinfull electrons (eg Hubbard model) 
Giamarchi et al.\cite{giamarchi} have shown that
the interaction, on the contrary, enhances the persistent current.
In this case, the increase of the current is closely related to the 
{\it decrease} of the spacial charge fluctuations or, 
equivalently, to the smoothing out in space of the charge density.
Indeed, for the repulsive Hubbard model,
the spin density wave (SDW) fluctuations are dominant in the ground state (GS).
On the other hand, for the attractive case, the spin fluctuations become 
gapped and thus pinning is favored. 
However, it should be noticed that, when 
$U \gg t$, the charge degrees of freedom 
can be mapped onto a spinless fermion model 
with $k_{f} \rightarrow 2k_{f}$ \cite{giamarchi}
so that the $4k_{f}$ component of the disorder is expected
to become important in this limit. 
At fixed $W$, we then expect a peak in the curve of D vs U.
ED data on small Hubbard rings \cite{Deng} have indeed shown
such a behavior. 
More recently, RG calculations taking
into account $4k_{f}$ scattering
have also confirmed this assumption \cite{MH}.
Second order perturbation calculations are also in good agreement
with the ED results \cite{Chiappe}.

Even though repulsive Hubbard interactions increase
the persistent currents, $4k_{f}$ scattering
prevents the observation, in a strictly 1D system, of 
amplitudes of the order of the experimental data. 
It should also be stressed  that models of spinless
fermions in the continuum lead to opposite results \cite{continuum}.
Indeed, it has been argued that, in the case of 1D continuous models, 
a repulsive interaction can increase the persistent current up to
values close to the 'clean' case.
These results emphasize the important role 
of both the spin and the nature of the interaction on the lattice.

\section{2D-Systems}

As stressed in the introduction, real experimental systems are 
3D anisotropic materials. Moreover, a diffusive regime can only be
realized in $d>1$.
A complete understanding of the persistent current
phenomenon then requires the study of multi-channel systems.
In section III, we shall present recent results
concerning the effect of the electronic interaction for both  
spinless fermion and Hubbard multi-channel models \cite{article3,article4}.
Note that we shall focus only on the {\it second} harmonic $I_{h2}$ of the 
current since (i) experimentally,
in the multi-ring experiments, the current was found to be 
$\frac{1}{2}$ periodic and (ii) theoretically, it is well understood that
the ensemble average (over filling or disorder)
suppresses the first harmonic of the current 
\cite{gilles1,gilles2,article3}.

Diagrammatic first order calculations (the spin is here irrelevant) 
have shown that the persistent currents are increased by the interactions 
\cite{Ambegaokar}.
More recently, Ramin et al.\cite{Ramin} have numerically shown by 
a Hartree-Fock (HF) approach, that the first order correction
of the persistent current was in agreement 
with the analytical treatment \cite{Ambegaokar}.
In both spinless or spinfull models,
the second harmonic is enhanced. 
However, a nearest neighbour interaction between spinless fermions 
tends to decrease the value of the {\it typical} current 
while a repulsive extended Hubbard interaction enhances it.

We have developped a self-consistent Hartree-Fock (SHF) method on 
finite lattices to treat the electron-electron 
interaction \cite{article3}. 
A systematic comparison with ED results on small lattices has revealed 
that this method was accurate even up to intermediate interaction strengths
\cite{article3}. While, in the small coupling limit, the SHF method 
reduces to the HF method, at larger coupling it takes into account
higher order powers of the interaction. 
Since the calculation is performed on (large) finite lattices 
quantum interferences due to the disorder potential are somehow 
treated exactly. It is important to note that this
method is different from the usual perturbative scheme \cite{Ambegaokar} 
where corrections to the current due to the electronic interaction are
calculated perturbatively.
Through the self-consistency relation, our procedure includes a 
resumation of higher order terms which becomes essential at moderate 
interaction strength. However, the direct relationship with a standard 
perturbative expansion remains unclear yet. 
Our study could be applied both to single or multi-ring experiments.
>From a theoretical point of vue, the difference relies simply 
in the absence or presence of particle number fluctuations.

\begin{figure}
\centerline{\epsfxsize 8cm \epsffile{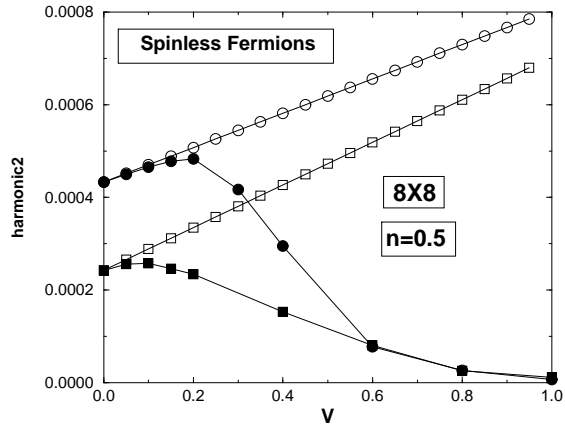}}
\caption{
$\big< I_{h2} \big>_{dis}$ as a function of V 
calculated within the SHF method for spinless fermions on 
a $8 \times 8$ cylinder. An average over 1000 disorder
configurations has been done.
Comparison between first order HF (open symbols) 
and SHF results (full symbols) at half-filling.
Circles correspond to $W=3t$ and squares to $W=4t$.}
\label{fig4}
\end{figure}

Section III is organized as follows: first, we compare, for
both spinless and Hubbard models, the first order correction 
(in the interaction) and the SHF correction to the second harmonic of the
persistent current.
As expected, the two methods are in very good agreement in 
the small interaction 
limit. For larger interaction strengths, higher order corrections 
(taken care of in the SHF method) become important for the case of 
the spinless fermion model while they remain small in the Hubbard case.
Secondly, we show that, as the interaction is switched on, 
the {\it decrease} (resp. {\it increase})
of the persistent current is always closely related to the {\it increase} 
(resp.{\it decrease}) of the charge density fluctuations.
Lastly, a finite size analysis of the data of the second
harmonic is performed. 

The starting hamiltonian reads:
\begin{eqnarray}
{\cal H} ={\cal H}_{K} + {\cal H}_{int} + {\cal H}_{des} .
\label{hamilt}
\end{eqnarray}
\noindent ${\cal H}_{K}$ is the usual kinetic part
containing the flux dependance, ${\cal H}_{des}$ is the term
due to the disorder,
\begin{eqnarray}
{\cal H}_{des}=\sum_{\bf i}\,w_{\bf i }\,n_{\bf i}.
\end{eqnarray}
The sites $\bf i$ are now chosen on a $L\times L$ lattice with periodic
boundary conditions in one direction. The disorder parameter W is restricted
to a finite interval for which the (non-interacting) system is as close as
possible to a diffusive regime \cite{mean_free_path}.
${\cal H}_{int}$ is the interacting part,
\noindent where, in the Hubbard case, $n_{\bf i}=n_{\bf i \uparrow}\,
+n_{\bf i \downarrow}$. In the spinless fermion case,
\begin{eqnarray}
{\cal H}_{int}^{S}= V\sum_{\bf i,\bf j}\,n_{\bf i}\,n_{\bf j},
\end{eqnarray}
\noindent where ${\bf i,j}$ stand for nearest neighbour sites 
and V is the strength of the screened interaction.
In the Hubbard case, the interaction part is local in space,
\begin{eqnarray}
{\cal H}_{int}^{H}= U\sum_{\bf i}\,n_{\bf i \uparrow}\,n_{\bf i \downarrow}.
\end{eqnarray}
\noindent
In the  SHF approximation, a mean-field type of decoupling is
performed,
\begin{eqnarray}
{\cal H}_{int}^{S} = 
& &-\sum_{\bf i,\bf j}
\delta t_{\bf i \bf j}\,c^{\dagger}_{\bf i}\,  c_{\bf j}\,+ 
\sum_{\bf i}\delta w_{\bf i}\,n_{\bf i} \nonumber \\
& & -\frac{1}{2}\,\sum_{\bf i,\bf j}\,V_{\bf ij}\, (\big< n_{\bf i}\big>
\big< n_{\bf j}\big>\,-\left|\,\big<c^{\dagger}_{\bf j}\,c_{\bf i}\big>\,\right|^{2})
\end{eqnarray}
\noindent which renormalizes the on-site disorder term (Hartree term) 
$\delta w_{\bf i}=\sum_{\bf j \ne \bf i }V_{\bf ij}\big< n_{\bf j}\big>$ and 
the hopping term (Fock term) 
$\delta t_{\bf i \bf j}=V_{\bf ij}
\big< c^{\dagger}_{\bf j}\,c_{\bf i}\big>$. 
The quantities $\big< n_{\bf j}\big>$ 
and $\big< c^{\dagger}_{\bf j}\,c_{\bf i}\big>$ 
have to be determined self-consistently.
Similarly, ${\cal H}_{int}^{H}$ becomes,
\begin{eqnarray}
{\cal H}_{int}^{H}=U\sum_{\bf i} ( \big< n_{\bf i \uparrow}\big>\,
 n_{\bf i \downarrow}+\big< n_{\bf i \downarrow}\big>\,n_{\bf i \uparrow}
-\big< n_{\bf i \downarrow}\big>\big< n_{\bf i \uparrow}\big> ) .
\end{eqnarray}
\noindent For simplicity the calculations have been done
in the paramagnetic sector (i.e.
$\big< n_{\bf i \downarrow}\big>=\big< n_{\bf i \uparrow}\big>$).
In this case the spin $\uparrow$ and $\downarrow$ are decoupled,
\begin{eqnarray}
{\cal H}_{int}^{H}=\sum _{\sigma}{\cal H}_{int}^{\sigma}
\end{eqnarray}
\noindent
where,
\begin{eqnarray}
{\cal H}_{int}^{\sigma}=U\sum_{\bf i} ( \big< n_{\bf i \sigma}\big> 
n_{\bf i \sigma}
-\frac{1}{2} \big< n_{\bf i \sigma}\big>^{2}) .
\end{eqnarray}
\noindent Note that, due to the disorder potential, the various mean-field
quantities are space dependent so that a numerical investigation on a
finite lattice is necessary. 
Similar equations also hold in the HF approximation but, in this case, the
expectation values are simply taken in the non-interacting GS.

\begin{figure}[bth]                                                        
\begin{center}                                                                
{\parbox[t]{6cm}{\epsfxsize 7.5cm                                              \epsffile{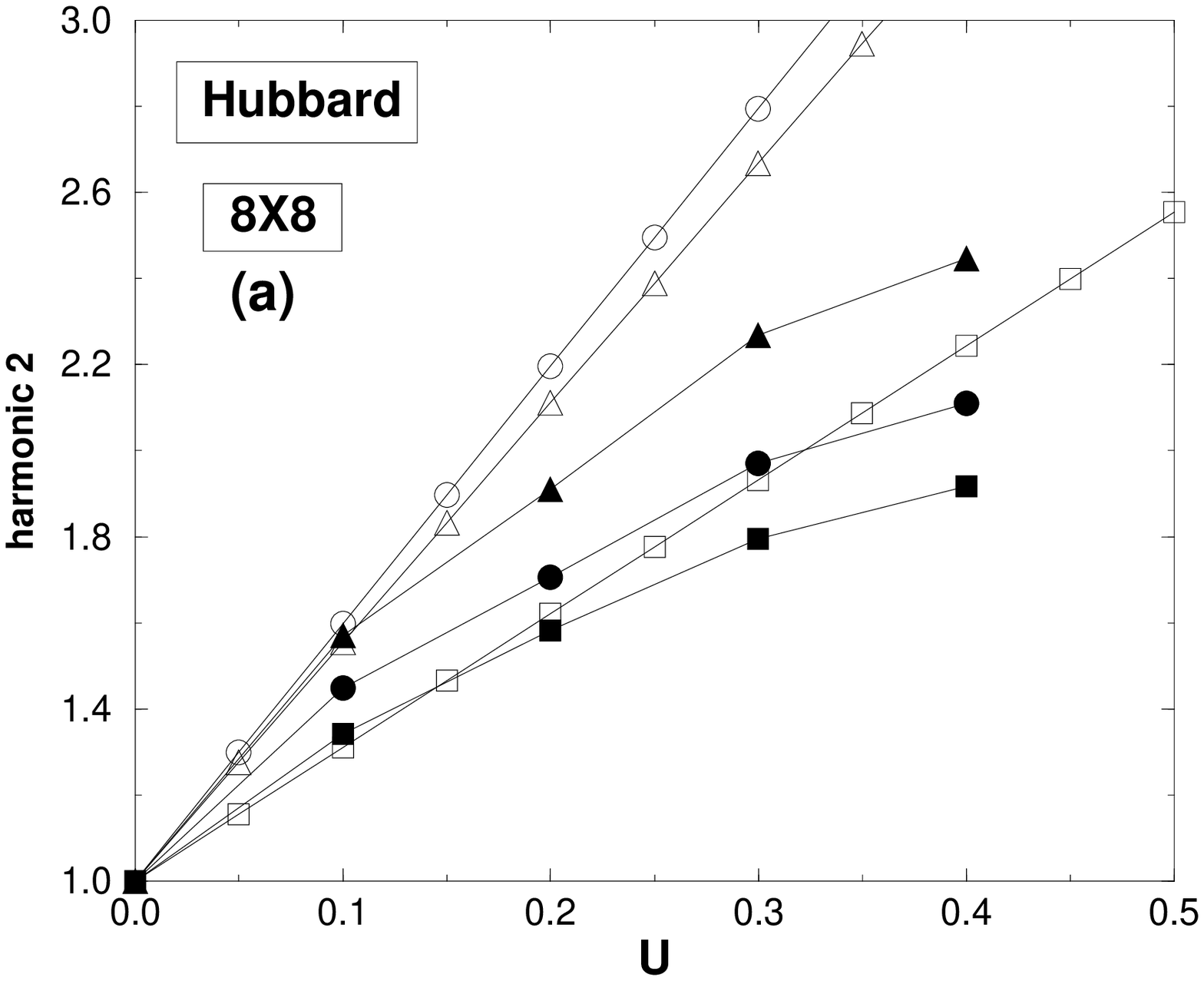} }                                                        \parbox[t]{6cm}{\epsfxsize 7.5cm                                              \epsffile{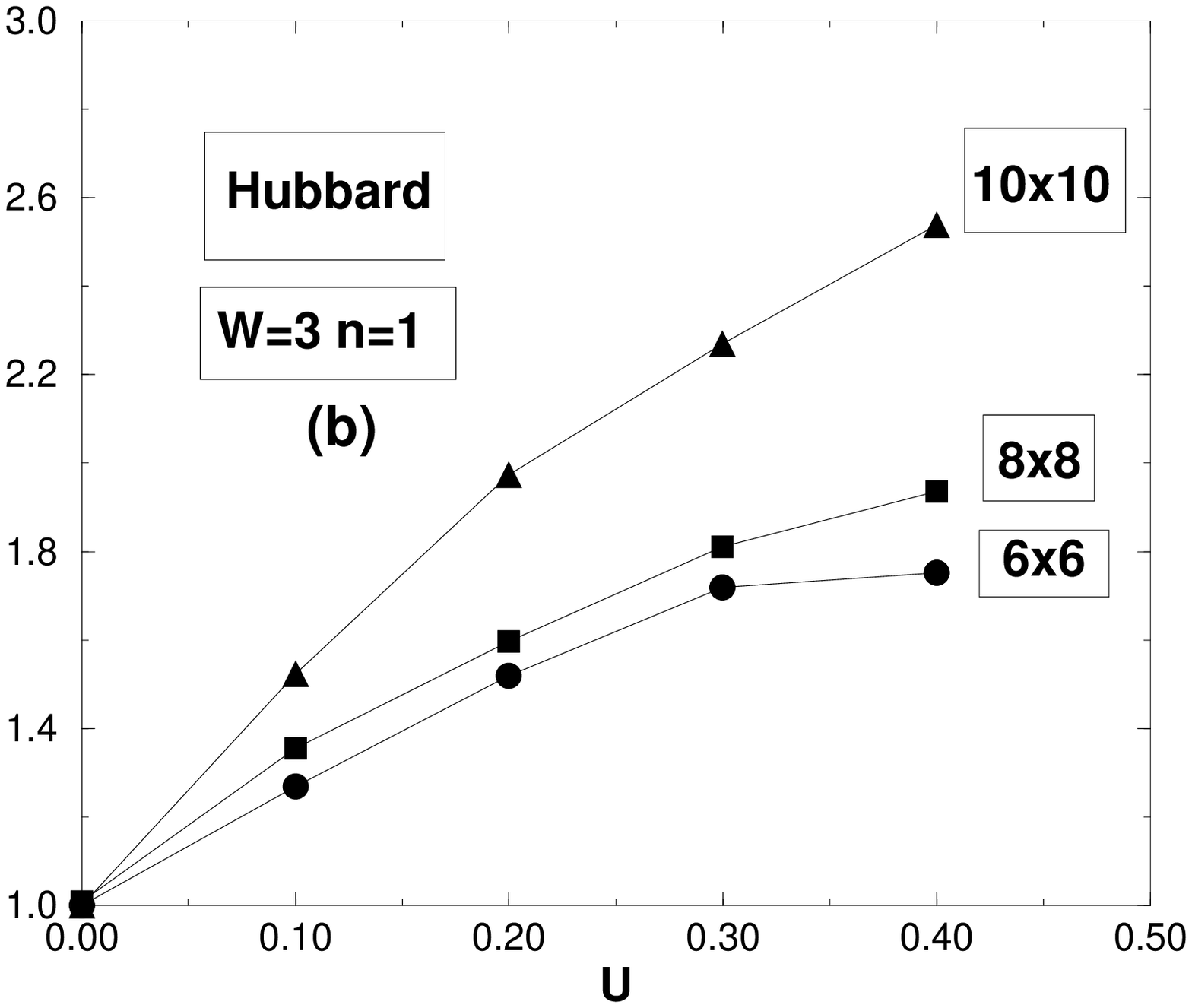}}                                                      
}   
\caption{(a)
(a) Ratio $\big< I_{h2} \big>_{dis}(U,W)/\big< I_{h2} \big>_{dis}(0,W)$ 
as a function of U calculated in the Hubbard model 
within both HF (open symbols) and SHF (full symbols) methods on 
a $8 \times 8$ cylinder. Circles correspond to $W=3t$ at quarter-filling,
squares to $W=3t$ at half-filling and triangles
to $W=4t$ at half-filling.
(b) Same as (a) at fixed $W=3t$ and at half-filling for several system 
sizes. 
}
\label{fig5}         
\end{center}
\end{figure}

Since the current is periodic of period 1 (in units of $\Phi_{0}$),
it can be expanded as a Fourier series,
\begin{eqnarray}
I(\Phi) = 
\sum_{\bf n}
I_{hn}\,sin(2\pi\,n\Phi), 
\label{Ih}
\end{eqnarray} 
where $I_{hn}$ are the harmonics of the current.
In fig.\ref{fig4} we have plotted $\big< I_{h2} \big>$ ($\big< \big>$
means average over disorder) at half-filling, for a $8 \times 8$ system,
as a function of V. $\big< I_{h2}\big>$ is calculated
by averaging over at least 1000 configurations of
the disorder.
As expected, we observe for weak values of the interaction
a perfect agreement between the SHF and the HF calculations.
However, as we increase V, the SHF results show a strong reduction 
of the current while the HF results predict an increase.
We also observe that the region of agreement between HF and SHF methods
is reduced as the disorder increases.
This means that, as the strength of the disorder increases,
higher order corrections in the interaction become significant
and, thus, a first order calculation is not sufficient.
We will see later on that this reduction is
related, as in the 1D case, to an increase of
the charge density fluctuations.
Using the SHF method, we have checked that this effect persists
for larger system sizes and away from half-filling.
In conclusion, the repulsive interaction are detrimental
to the persistent currents in spinless fermion models with
short range interactions.
Note however that, in a different regime, for much {\it stronger} disorder, 
an enhancement of the persistent currents by the interactions has been 
found \cite{Berkovits}.

However, a different behavior is observed in the case of
the Hubbard model as seen in fig.\ref{fig5}, showing the relative effect of 
the interaction on the current, 
i.e. $\frac{\big< I_{h2}\big> (U,W)}{\big< I_{h2}\big> (0,W)}$.
Fig.\ref{fig5}(a) reveals that both SHF and HF calculations 
predict an increase of the current and that
the first order calculation gives relatively good results.
It is interesting to note that the first order calculations always give
higher values of the currents.
In Fig.\ref{fig5}(b), we have plotted the same quantity, 
for a fixed density, but for different system sizes.
It clearly indicates that, as the size of the system
increases, the effect of U becomes stronger.

\begin{figure}[bth]                                                        
\begin{center}                                                                
{\parbox[t]{6cm}{\epsfxsize 7.5cm                                              \epsffile{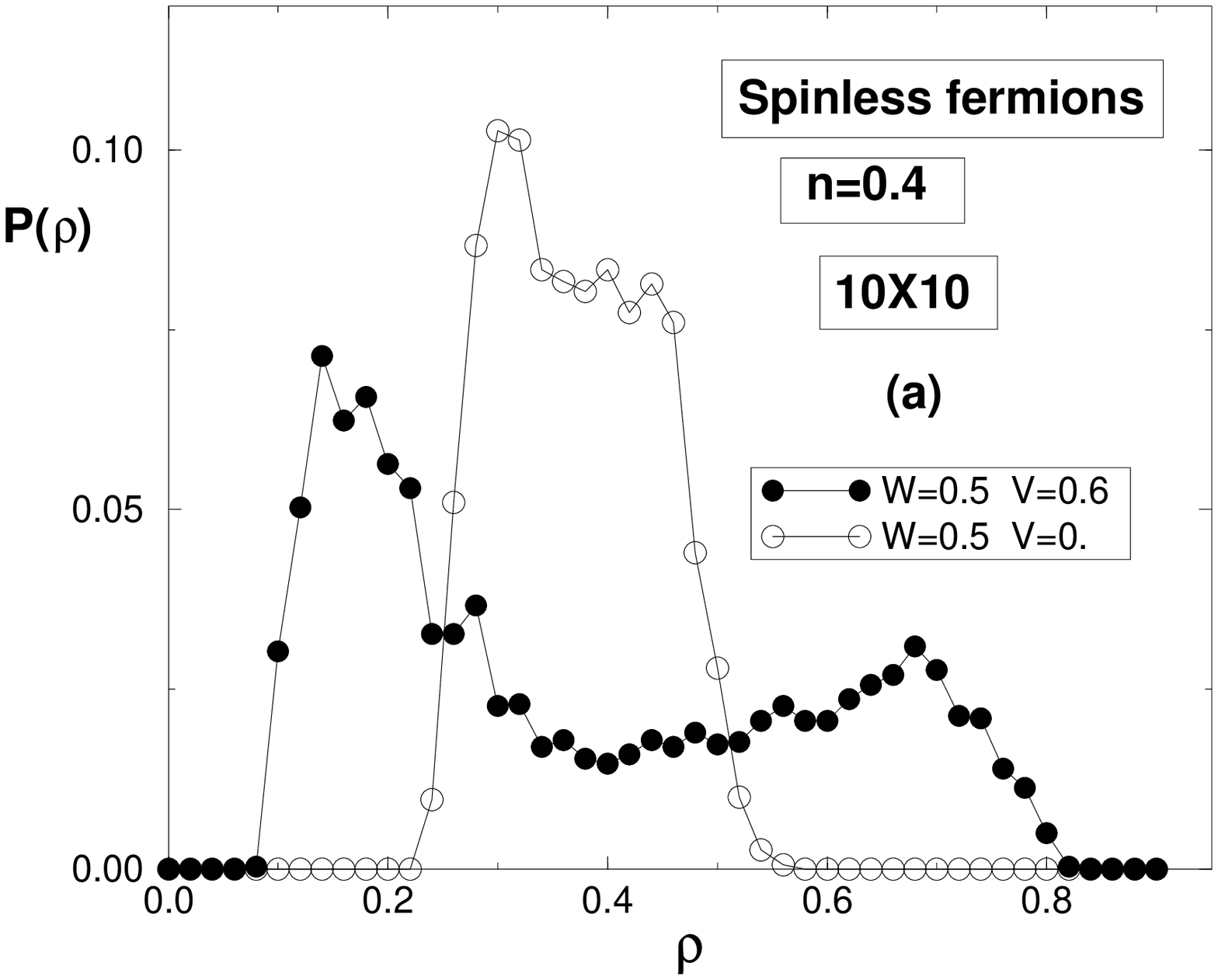} }                                                        \parbox[t]{6cm}{\epsfxsize 7.5cm                                              \epsffile{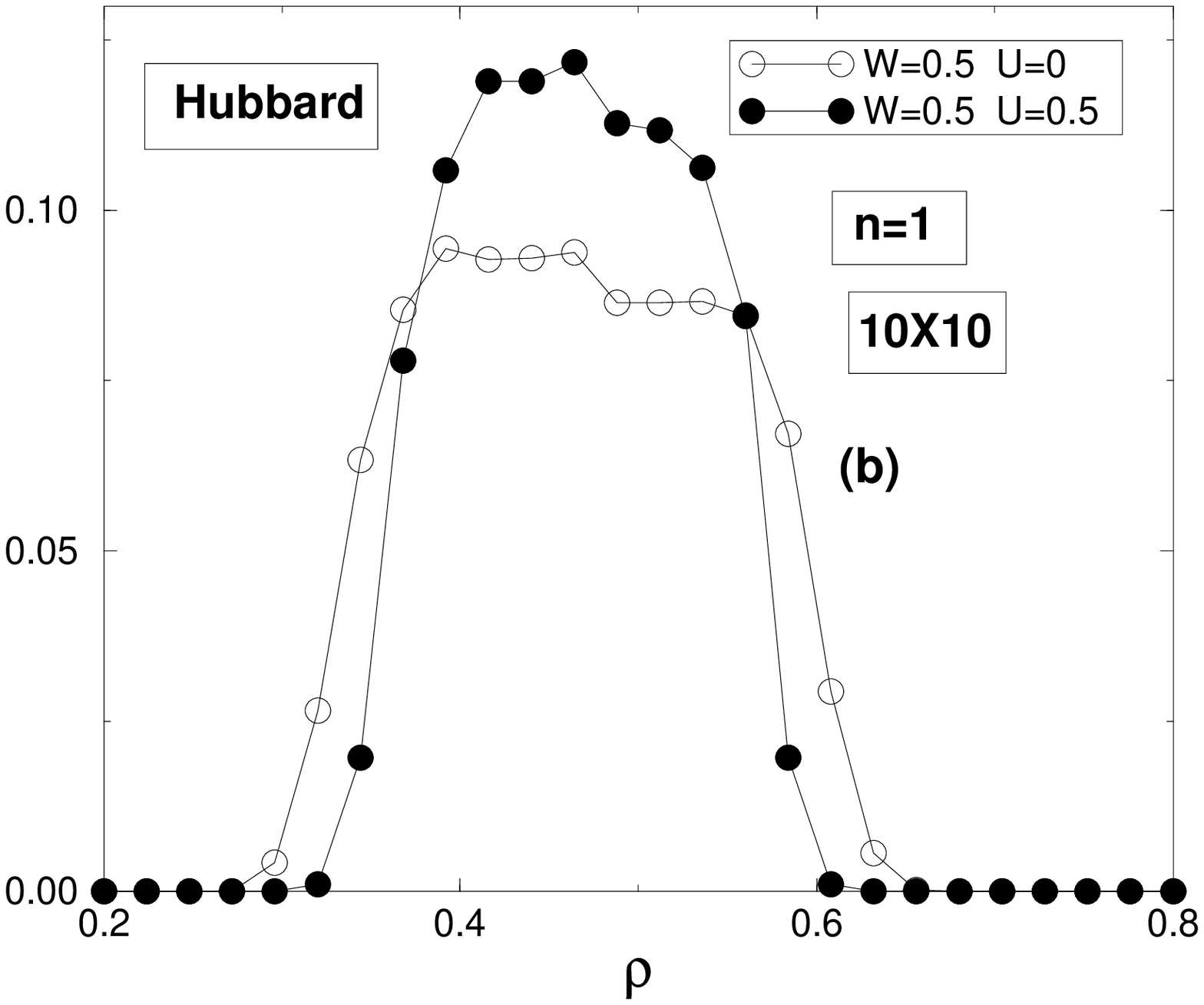}}                                                      
}   
\caption{Distribution $P(\rho)$ in both spinless
fermion (a) and Hubbard (b) models, for a $10 \times 10$ cylinder,
and different band fillings n.
The values of V, W and U are indicated in the figures.
An average over 30 configurations of the disorder has been performed.
}
\label{fig6}         
\end{center}
\end{figure}

At this point, these first results suggest the importance of the 
nature of the interaction. Indeed, in the spinless case, the currents
are strongly reduced by the interaction while, on the contrary,
the Hubbard repulsive interaction enhances the currents.
In addition, it has been shown that in the spinless fermion case,  
higher order terms become rapidly dominant even for relatively modest
values of the interaction strength so that
a simple HF approach is not sufficient.

Let us now try to build up a physical picture of this phenomenon.
The distribution of the local charge density $n_{\bf i}$ 
turns out to be directly connected to the localization.
Considering many configurations of disorder (labeled by some integer k), 
we assume that the related local site densities $n_{\bf i}^{k}$ (where
the subscript k stands for the disorder configuration) are 
independent realizations of a statistical variable $\rho$.
In fig.\ref{fig6} we have plotted the distribution of 
the local density $\rho$ for spinless fermion (a)
and Hubbard (b) models.
We clearly observe that the shape of the distribution 
changes with the interaction.
In the spinless fermion case, two peaks
clearly appear. This suggests the existence of two categories of sites with
rather large and small charge densities, respectively.  
On the contrary, in the Hubbard case (b), we observe that
the interaction has now the opposite effect i.e. the distribution shrinks
around the average value. In other words, the interaction tends to 
homogenize the charge density in space. 

\begin{figure}[bth]                                                        
\begin{center}                                                                
{\parbox[t]{6cm}{\epsfxsize 7.cm                                              \epsffile{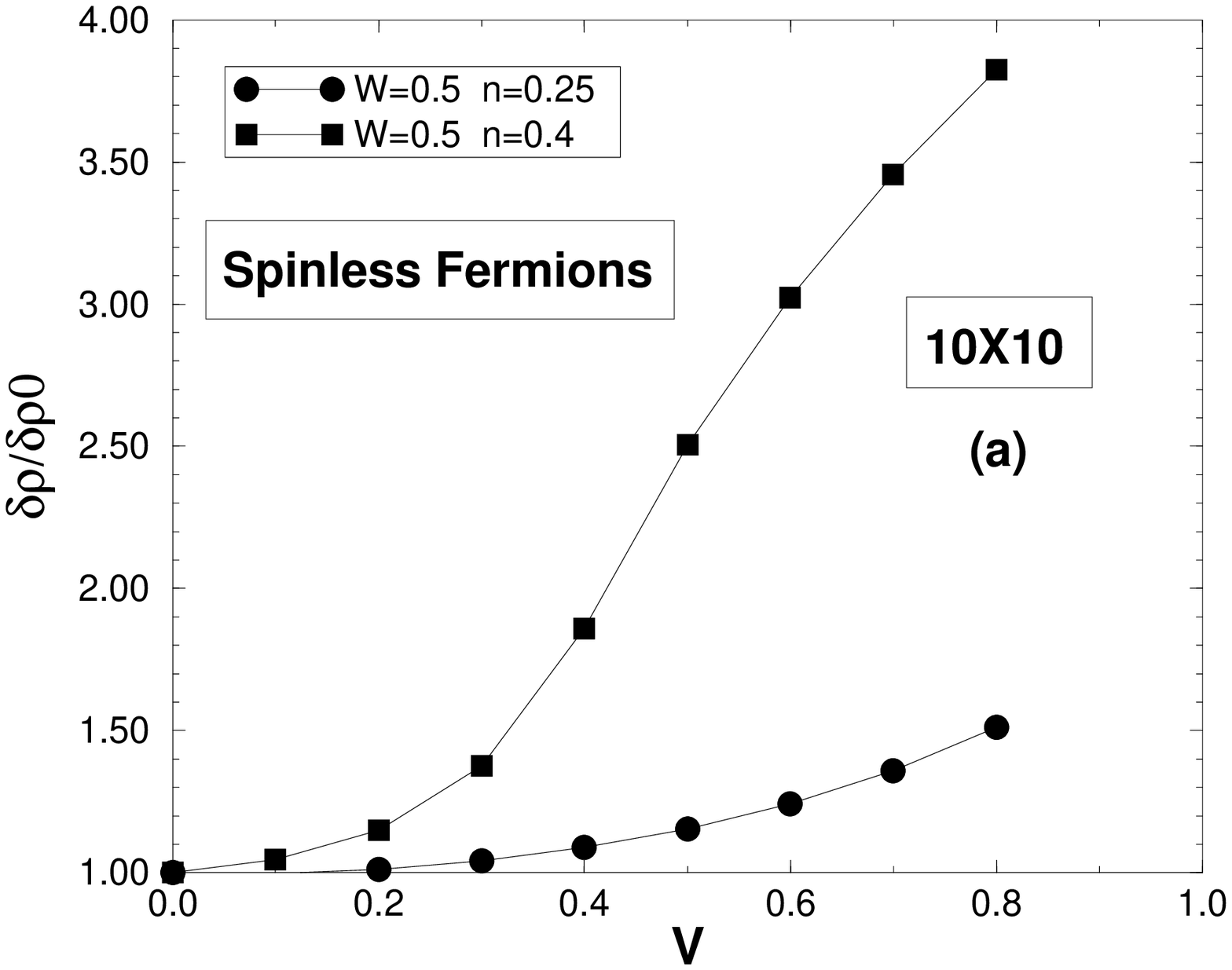} }                                                        \parbox[t]{6cm}{\epsfxsize 7.cm                                              \epsffile{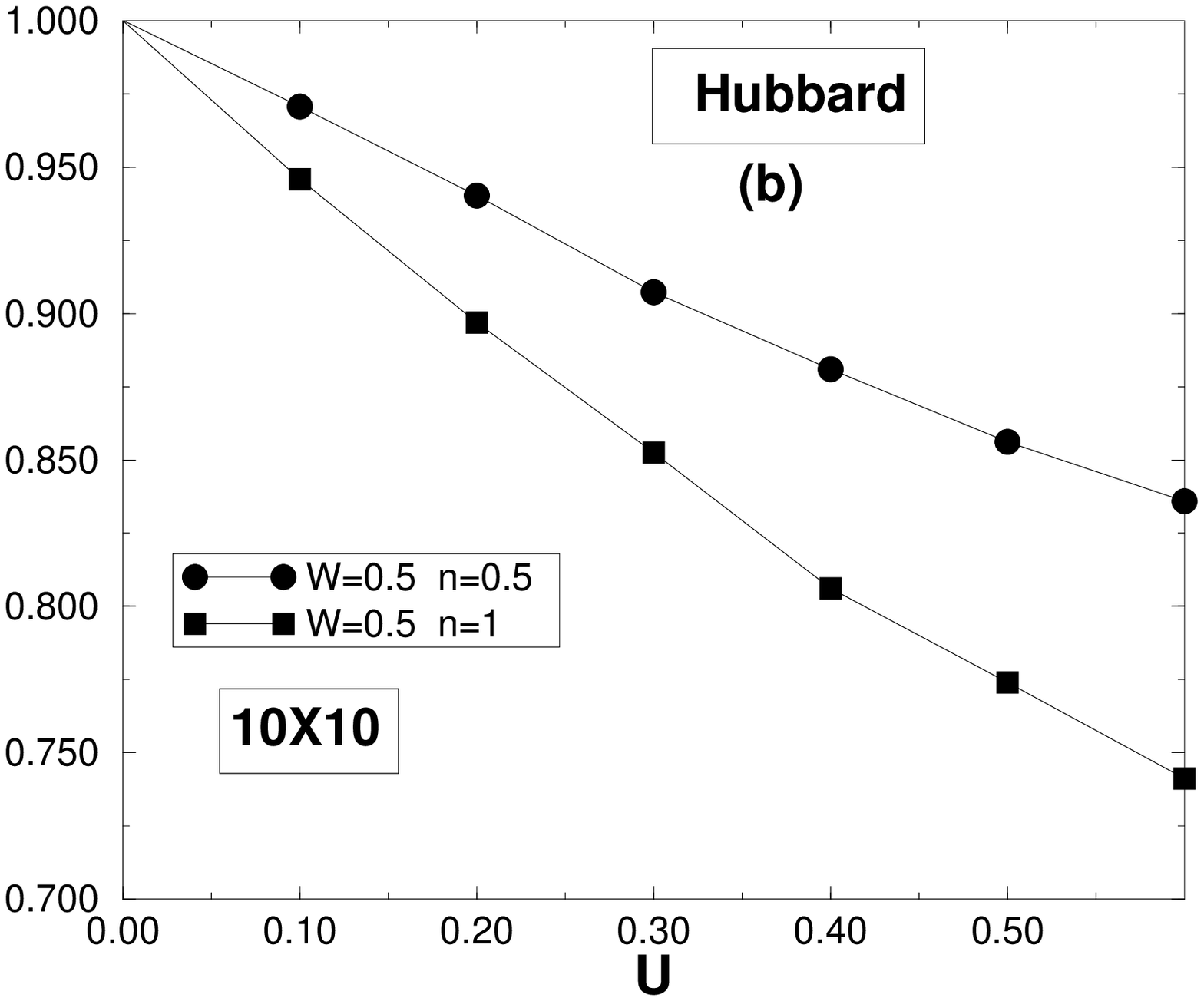}}                                                      
}   
\caption{
Ratio $\delta \rho (VU,W)/\delta \rho (0,W)$,
(VU is V or U) as a function
of the interaction parameter (U or V) for different filling factors n.
The calculations have been done on a $10 \times 10$ cylinder
with 30 configurations of the disorder.
}
\label{fig7}         
\end{center}
\end{figure}

Let us now turn to more qualitative
results. For this purpose, let us define,
\begin{eqnarray}
\delta \rho_{k}=\sqrt{\frac{1}{L^{2}} \sum_{i=1}^{L^{2}}(
\big< n_{i}^{k}\big> - n )^{2}}
\end{eqnarray}
and,
\begin{eqnarray}
\delta \rho=\frac{1}{N_{cdes}} \sum_{k=1}^{N_{cdes}}\delta \rho_{k}
\end{eqnarray}
where L is the length of the system, $N_{cdes}$ is
the number of disorder configurations, and
$\big< n_{i}^{k}\big>$ are calculated self-consistently.
We expect, for big systems, 
$\delta \rho_{k}$ to become independent of the disorder
configuration. Results for $\delta \rho$ vs U or V 
are plotted in fig.\ref{fig7} for various fillings.
For spinless fermions, it appears that $\delta \rho$
increases with V at any fillings.
For example, for $n=0.4$, the width increases by almost a factor 4 at
$\frac{V}{t}=0.8$.
However, in the Hubbard case, the effect of U is just the opposite e.g.
the reduction of $\delta \rho(U,W)$ is
larger than 25\% at half filling.
Note that, in general, the magnitude of the effect (relative increase 
or decrease with the interaction) increases with the density,
as commensurability is approached. 

\begin{figure}[bth]                                                        
\begin{center}                                                                
{\parbox[t]{6cm}{\epsfxsize 7.5cm                                              \epsffile{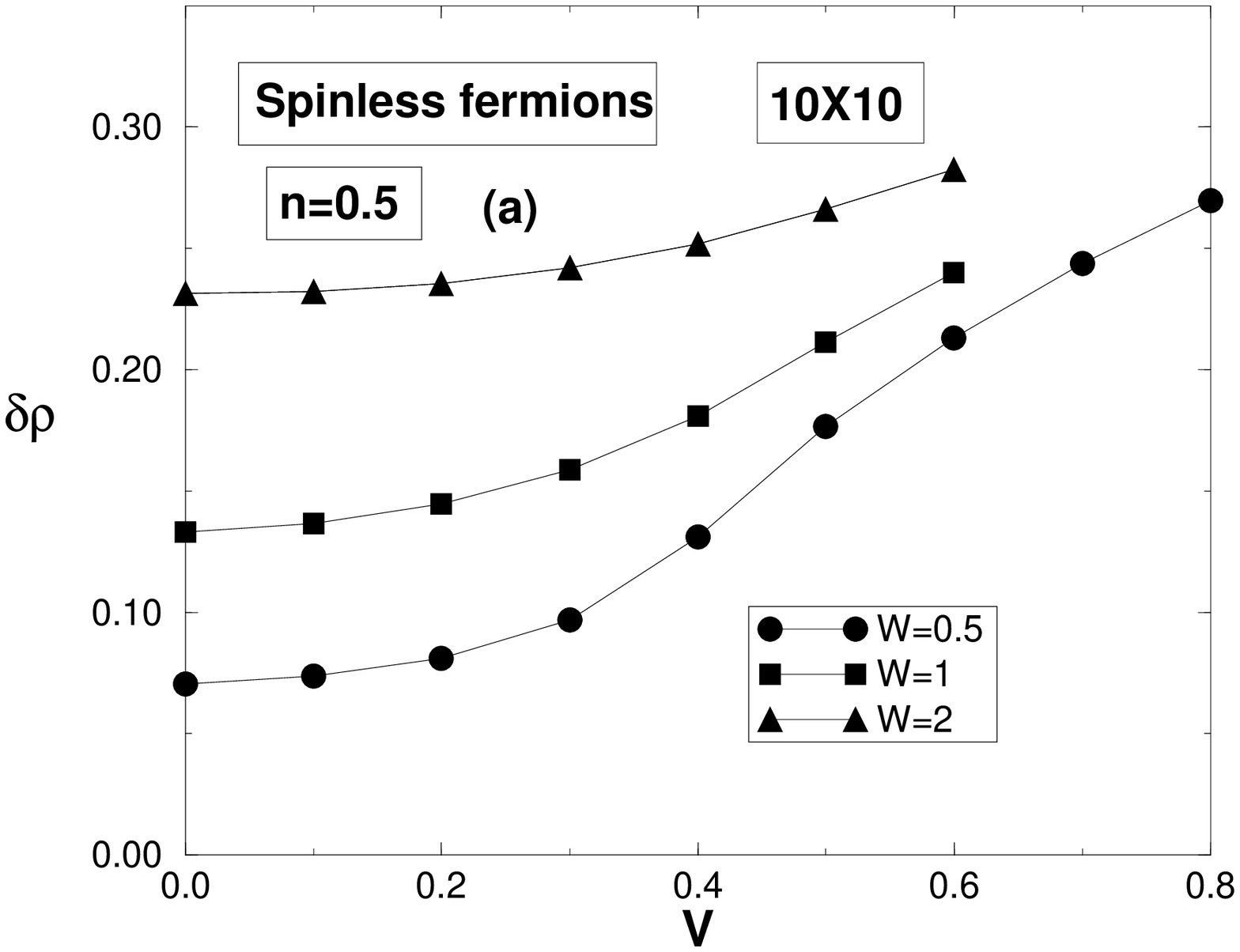} }                                                        \parbox[t]{6cm}{\epsfxsize 7.5cm                                              \epsffile{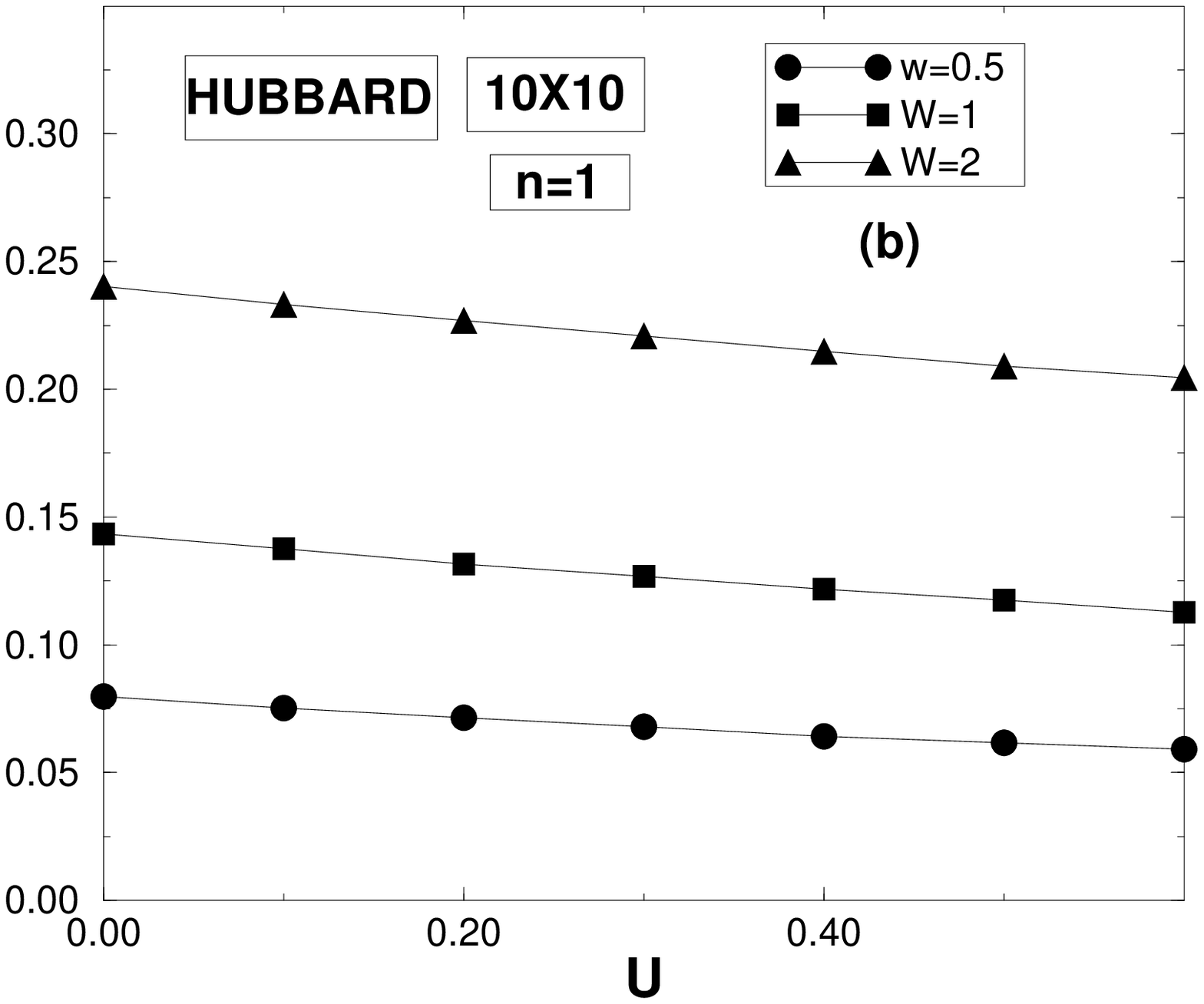}}                                                      
}   
\caption{$\delta \rho (V or U,W)$ as a function
of the interaction parameter (V or U) for different 
disorder $W/t$. 
The calculations have been done on a $10 \times 10$ cylinder
with 30 configurations of the disorder.
}
\label{fig8}         
\end{center}
\end{figure}

To finish, we now turn to dependence of the width of the 
distribution $\delta \rho $ on the disorder parameter W
summarized in Fig.\ref{fig8}.
As expected, for vanishing interaction strength, the distribution of the
local densities increases almost proportionnaly with W.
For finite interaction, Fig.\ref{fig8} shows that the {\it sign} of the 
effect observed depends 
only on the nature of the interaction (U or V) and not on the disorder. 
The magnitude of the increase/decrease seems to depend roughly on the 
combination $\frac{V}{W}$ or $\frac{U}{W}$.

\section{Conclusion}
 
In conclusion, we have presented data beyond first
order perturbative calculations. In 1D, exact diagonalisation data 
clearly show that the electronic interaction enhances localization.
In 2D, a self-consistent mean-field treatment of the interaction
is used for the first time in the case of disordered fermions on a
lattice. The SHF method has proven itself to be a good tool to study 
the effect of the interaction for $d>1$.
It provides strong evidences 
that spinless and spinfull fermions behave differently. 
The above SHF results show that the increase 
of the persistent current is strongly connected to the spin degrees of freedom.
Although the calculations presented here concern 2D systems, we expect that,
for bigger systems and higher connectivity, the pinning 
with the impurities becomes weaker. 
However, quantitative data are still missing for realistic mesoscopic sizes
and more work is certainly needed to know whether such models 
can account for the correct magnitude of the experimentally observed
current.

D.P. acknowledges support from the
EEC Human Capital and Mobility Program under grant CHRX-CT93-0332.
{\it Laboratoire de Physique Quantique (Toulouse)}
is {\it Unit\'e Mixte de Recherche No. UMR5626 du CNRS}.
We thank IDRIS (Orsay, France) for generously providing CPU time on the
C94 and C98 supercomputors.

\vspace{2cm}
\begin{center}
\small{\bf COURANTS PERSISTENTS 
DANS LES SYST\`EMES DE FERMIONS CORR\'EL\'ES}
\bigskip
\end{center}
\bigskip

{\small 
Les courants persistents dans des anneaux m\'esoscopiques d\'esordonn\'es
sont \'etudi\'es \`a l'aide de m\'ethodes de diagonalisations exactes 
\`a 1D et de m\'ethodes Hartree-Fock auto-coh\'erentes \`a 2D. Dans le cas 
unidimensionnel d'un mod\`ele de fermions sans spin, les interactions 
\'electroniques augmentent toujours la localisation. En revanche, dans le 
cas de syt\`emes bi-dimensionnels de fermions avec spin 1/2, le courant 
permanent est nettement augment\'e par les corr\'elations \'electroniques.
Il est montr\'e que ce ph\'enom\`ene est intrins\`equement li\'e \`a 
la diminution des inhomog\'en\'eit\'es de densit\'e de charge dans l'espace
provoqu\'ee par les interactions \'electroniques. Cette \'etude d\'emontre 
l'importance cruciale des degr\'es de libert\'e de spin. 
}
\bigskip

\end{document}